\documentclass[11pt]{article}
\usepackage[textwidth=15.2cm,textheight=22cm]{geometry}
\usepackage{amsmath,amssymb}
\usepackage{latexsym}
\usepackage{multicol}
\usepackage{graphicx}
\usepackage{bm}
\usepackage{tikz}
\usepackage{cite}
\usepackage{tikz}
\usetikzlibrary{trees}
\usepackage{varwidth}
\usepackage{pifont}
\usepackage{cancel}
\usepackage{centernot}
\usepackage{mathtools}
\usepackage{float}
\usepackage{simpler-wick}
\usepackage{xcolor}

\newcommand{\xdownarrow}[1]{%
  {\left\downarrow\vbox to #1{}\right.\kern-\nulldelimiterspace}
}
\allowdisplaybreaks[1]

\newcommand{\del}{\partial}
\newcommand{\be}{\begin{equation}}
\newcommand{\ee}{\end{equation}}
\newcommand{\ba}{\begin{eqnarray}}
\newcommand{\ea}{\end{eqnarray}}

\newcommand{\rom}[1]{\uppercase\expandafter{\romannumeral #1\relax}}

\def\ba{\bar A}

\def\beq{\begin{equation}}
\def\eeq{\end{equation}}

\newcommand{\nn}{\nonumber}

\newcommand{\ndt}{\noindent}

\def\bea{\begin{eqnarray}}
\def\eea{\end{eqnarray}}
\def\beas{\begin{eqnarray*}}
\def\eeas{\end{eqnarray*}}
\def\sla{\raise.15ex\hbox{$/$}\kern-.57em}

\def\parm{{\partial}_{-}}

\def\m#1{\mathcal#1}

\def\spa#1.#2{\left\langle#1\,#2\right\rangle}
\def\spb#1.#2{\left[#1\,#2\right]}

\def\m{\mu}

\def\pa{{\partial}}

\date{}
\begin{document}
\begin{titlepage}
\begin{flushright}    
{\small $\,$}
\end{flushright}
\vskip 1cm
\vskip 1cm
\centerline{\Large{\bf{Nicolai maps and uniqueness in the light-cone gauge}}}
\vskip 0.3 cm
\centerline{Nipun Bhave and Saurabh Pant}
\vskip 0.3cm
\centerline{\it {Indian Institute of Science Education and Research}}
\centerline{\it {Pune 411008, India}}
\vskip 1.5cm
\centerline{\bf {Abstract}}
\vskip .4cm
\ndt We compute the Nicolai map for the supersymmetric Yang-Mills theory, in the light-cone gauge, to the second order in the coupling constant for all critical dimensions ($d=3,4,6,10$). The process of integrating out unphysical degrees of freedom in this gauge, produces a four fermion interaction term. We show that, to the order investigated here, this term is harmless.  We demonstrate the existence of a particularly `simple'  map in $d=4$ in the light-cone gauge and address the issue of uniqueness in the context of the map. We also investigate the map in the light-cone superspace in $d=4$. 

\vskip .5cm
\ndt 
\vfill
\end{titlepage}

\section{Introduction}

\ndt Supersymmetric gauge theories have been studied extensively because of their interesting ultraviolet properties. The flagship theory, the maximally supersymmetric $\mathcal{N}=4$ Yang-Mills theory, for example, is a perturbatively finite quantum theory in four dimensions. An alternative perspective on supersymmetric gauge theories is offered by the Nicolai map~\cite{Nicolai:1979nr,Nicolai:1980jc,Nicolai:1984jg} which makes the formulation of these theories possible without the use of anti-commuting variables~\cite{Lechtenfeld:2023gcq,Lechtenfeld:1984me}. \\

\ndt While this map has been investigated in the Landau gauge~\cite{Ananth:2020gkt,Ananth:2020lup,Ananth:2020jdr}, its study in other gauges has been limited~\cite{Malcha:2021ess,Lechtenfeld:2021yjb,deAlfaro:1985xz}. Not surprisingly, the map appears more complicated in the axial gauge (and in the light-cone gauge) than in the Landau gauge~\cite{Malcha:2021ess}. In this paper, we work with the $LC_2$ light-cone gauge approach which means that only the physical degrees of freedom of the theory are retained. This means that the role of the little group is explicit while manifest covariance is sacrificed. The close link between the light-cone gauge and spinor-helicity variables~\cite{Ananth:2012un}, implies that studying the map in this language could prove useful in the study of scattering amplitudes. \\ 

\ndt One aim of this paper is to demonstrate the existence of a particularly `simple'  map in $d=4$ in the light-cone gauge and consequently in terms of the helicity variables. We also find a map that works in $d=3,4,6$ and $10$. When we write this map explicitly in $d=4$ in terms of the helicity variables, we find that this map is distinct from the `simple' $d=4$ map. This raises the question of the uniqueness of the map which we attempt to address in this paper~\cite{Ananth:2020jdr,Lechtenfeld:2022qpa }.\\

\ndt The second part of this paper focuses on superspace. The first all-order proof of finiteness for the $\mathcal{N}=4$ Yang-Mills theory was provided using light-cone superspace~\cite{Mandelstam:1982cb,Brink:1982wv}. It is therefore of interest to ask how this proof of finiteness may be achieved within the framework of the Nicolai map. With this goal in mind, we find a non-linear and non-local transformation for the $\mathcal{N}=1$ superfield with a trivial Jacobian such that the full $\mathcal{N}=1$ Lagrangian is mapped to a free Lagrangian in light-cone superspace. \\
\\
\ndt A future direction of interest would be to ask whether the issues addressed in this paper could be extended to supergravity formulated in light-cone superspace. 
 \\
\\
\ndt The paper is organized as follows. In section 2, we start with the $\mathcal{N}=1$ Yang-Mills theory in the light-cone gauge to write a map in $d=4$. In section 3, we generalize the map to all critical dimensions ($d=3,4,6,10$). We also establish its connection with the map derived in general gauges~\cite{Malcha:2021ess}. In section 4, we comment on the uniqueness of the light-cone Nicolai map in $d=4$. In section 5, we compute the Nicolai map in light-cone superspace in four dimensions. In the last section, we discuss the possible connection between the Nicolai map and the quadratic form structure in the light-cone Hamiltonian for pure Yang-Mills theory. 
\vskip 0.3 cm
\ndt {\bf Note}: While working on the manuscript, we became aware of \cite{Lechtenfeld:2024uhi}, which contains some overlapping results.\\

\section{Light-cone Nicolai map in $d=4$}

\ndt We work with the light-cone coordinates given by
\bea
x^{\pm}=\frac{x^0\,\pm \,x^3}{\sqrt{2}}\,,\hspace{1cm}  x=\frac{x^1+ix^2}{\sqrt{2}}\,,\hspace{1cm}\bar{x}=x^*\,,
\eea

\ndt and their derivatives $\del_{\pm}\,(-\del^{\mp})\,,\,\bar{\del}\,,\,\del$ respectively$\,$\footnote{We define the `inverse' derivative using the step function: $\frac{1}{\del^+}\,f(x^-)\,\equiv\,-\int dy^-\,\,\theta(x^--y^-)f(y^-)$.}.\\

\ndt We start with the $\mathcal{N}=1$ supersymmetric Yang-Mills Lagrangian in the light-cone gauge written purely in terms of the physical degrees of freedom~\cite{Brink:1982pd}. Details of the procedure to obtain this Lagrangian are presented in Appendix A. The Lagrangian written entirely in terms of physical fields - the gauge fields and fermion fields ($A,\bar{A},\chi,\bar{\chi}$) - is
\bea
\label{Final L}
\hspace{1 cm}\mathcal{L}&=&\bar{A}^a \Box A^a - 2g f^{abc} \left(\frac{\bar{\partial}}{\partial^+}A^a\partial^+\bar{A}^b A^c+\frac{\partial}{\partial^+}\bar{A}^a\partial^+A^b\bar{A}^c\right) \nn\\
&& \hspace{1 cm} -2g^2 f^{abc} f^{ade} \frac{1}{\partial^+}\left(\partial^+A^b\bar{A}^c\right)\frac{1}{\partial^+}\left(\partial^+\bar{A}^d A^e\right) \nn\\
&&\hspace{-1.4 cm}+\,\frac{i}{\sqrt{2}}\bar{\chi}^a\left(\frac{\Box}{\del^+}\delta^{ac}-2gf^{abc}\frac{1}{\del^+}(\partial\bar{A^b}+\bar{\partial}A^b) +2gf^{abc}{\bar{A}^b}\frac{\partial}{\del^+}\right)\chi^c + i\sqrt{2} gf^{abc} \bar{\chi}^a\frac{\bar{\partial}}{\del^+}(A^b \chi^c)\nn\\
&&\hspace{ -0.5 cm}\,+\,i\sqrt{2}\,g^2 f^{abc}f^{bde}\,\bar{\chi}^a\frac{1}{\del^{+2}}(A^d\del^+\bar{A}^e+\bar{A}^d\del^+{A}^e)\chi^c-i\sqrt{2}\,g^2 f^{abd}f^{bec}\bar{\chi}^a \bar{A}^d\frac{1}{\del^+}(A^e\chi^c)\nn\\
&&\hspace{1cm}+g^2f^{abc}f^{ade}\frac{1}{\del^+}(\bar{\chi}^b\chi^c)\frac{1}{\del^+}(\bar{\chi}^d\chi^e)\,.\nn\\
\eea

\ndt where $\Box=(-2\del^+\del^-+2\del\bar{\del})$ and the $f^{abc}$ are the structure constants of the gauge group.\\

\ndt The statement of the Nicolai map is the following: there exists a non-linear and non-local transformation $\mathcal{T}_g(A)$ which satisfies the following three properties:\\

\ndt 1. The transformation $\mathcal{T}_g(A)$ when substituted in the free bosonic Lagrangian (Maxwell theory) yields the full interacting bosonic Lagrangian (Yang-Mills theory).\\

\ndt 2. The Jacobian of the transformation is equal to the fermion determinant (or the product of the fermion and ghost determinants in cases where the gauge choice does not eliminate all unphysical degrees of freedom).\\

\ndt 3. The transformation preserves the gauge choice.\\

\ndt Essentially, this means that one works with a free bosonic theory to compute correlators in a supersymmetric gauge theory - through the inverse transformations $\mathcal{T}_g^{-1}(A^{'})$~\cite{Nicolai:2020tgo}. 
\\
\subsection{The transformation}
\label{Trans}
\ndt We now write down a field transformation for the physical fields $A^a$ and $\bar{A}^a$ by trial and error so the Yang-Mills Lagrangian may be written as a purely kinetic term in the new (primed) variables: $\bar{A}^{'\,a}\Box A^{'\,a}$. We introduce a Green's function through $\Box C(x-y)=-\delta^{(4)}(x-y)$ to write such an ansatz upto $\mathcal{O}(g^2)$
\bea
\label{Tnmap}
A^{'\,a}(x,g;A,\bar{A})&=&A^a(x)+2gf^{abc} \int dy \,\,\del^+C(x-y) \frac{\bar{\partial}}{\del^+}\,A^b(y)A^c(y)\nn\\
&&-g^2 f^{abc}f^{bde} \int dy \,\,\del^+ C(x-y) A^c(y) \frac{1}{\del^{+\,2}}\left(\del^+A^d(y)\bar{A}^e(y)\right)\nn\\
&&-\,2g^2 f^{abc}f^{bde}  \int dy\, dz\,\, \left(\del\,C(x-y) \bar{A}^c(y) - \del^+ C(x-y)\frac{\partial}{\del^+} \bar{A}^c(y) \right)\nn\\
&& \hspace{4cm}\times\,\,{{\del^+}} \,C(y-z)  \frac{\bar{\partial}}{\del^+} A^d(z)A^e(z)\ .
\eea
\ndt Here $dy\,,\,dz$ denote the four dimensional space-time measure. In this section, all measures and delta functions will be assumed to be four dimensional and the dimension will be suppressed henceforth. The transformation for $\bar{A}^{'\,a}$ is just the complex conjugate of the above.  \\

\ndt In a covariant approach, the map at order $g^2$, contains terms of the form $\del\, C\, A \,\del\, C \,A \,A$ (with space-time indices and color indices suppressed), all of which contribute to the Jacobi determinant. In the light-cone Nicolai map at order $g^2$, we find that there is a term with a single Green's function (line 2 of eq. (\ref{Tnmap})). This term produces the pure Yang-Mills quartic vertex (line 2 of eq. (\ref{Final L})) but does not contribute to the Jacobian at order $g^2$ as we show below.\\

\ndt The functional variation of the fields are 
\bea
\frac{\delta A^{a}(x)}{\delta A^{b}(w)}=\frac{\delta \bar{A}^{a}(x)}{\delta \bar{A}^{b}(w)}=\delta^{ab}\delta(x-w)\,,\;\;\;\;\;\frac{\delta A^{a}(x)}{\delta \bar{A}^{b}(w)}=\frac{\delta \bar{A}^{a}(x)}{\delta A^{b}(w)}=0\,.
\eea

\ndt The Jacobi matrix\footnote{The matrix elements $\cfrac{\delta A^{'\,a}(x)}{\delta\bar{A}^m(w)}$ and $\cfrac{\delta \bar{A}^{'\,a}(x)}{\delta A^m(w)}$ do not contribute to the trace at order $g^2$} of the above transformation (\ref{Tnmap}) is

\bea
\frac{\delta A^{'\,a}(x)}{\delta A^m(w)}&=&\delta^{am}\delta(x-w)+2gf^{abc} \int dy \,\,\biggl \{\del^+C(x-y) \frac{\bar{\partial}}{\del^+}\,\delta^{bm}\delta(y-w)A^c(y) \nn\\
&&\hspace{3cm}+\,\del^+C(x-y) \frac{\bar{\partial}}{\del^+}\,A^b(y)\,\delta^{cm}\delta(y-w)\biggl \}\nn\\
&&-2g^2 f^{abc}f^{bde} \int dy\, dz\,\biggl( \del\,C(x-y) \bar{A}^c(y) - \del^+ C(x-y)\frac{\partial}{\del^+} \bar{A}^c(y) \biggr) \nn\\
&&\times\,\,\biggl \{ {{\del^+}} \,C(y-z) \frac{\bar{\partial}}{\del^+}\delta^{dm} \delta(z-w)A^e(z) + {{\del^+}} \,C(y-z)  \frac{\bar{\partial}}{\del^+} A^d(z)\,\delta^{em} \delta(z-w)\biggl \}\,,\nn\\
\eea
\ndt where we have dropped all terms that vanish after taking the trace as they are proportional to $\del_{\mu} C(0)$. The Jacobi determinant of the map can be computed using the relation 
\bea
\label{Identity 1}
\log \det(1+{\bf X})={\rm Tr} \log(1+{\bf X}) ={\rm Tr} \,{\bf X}- \frac{1}{2}{\rm Tr}\, {\bf X}^2 \pm..
\eea

\ndt After partial integrations, taking the trace (by setting $a=m$ and $x=w$), using $f^{abc}f^{abd} = n\delta^{cd}$ and integrating over $x$, the Jacobi determinant upto $O(g^2)$ reads
\bea
\label{Jacobian}
\hspace{-1 cm} \log \det\left(\frac{\delta A_i'^a(x)}{\delta A_j^m(w) }\right)&=&2ng^2 \int dx\, dy\,\,{\biggl \{} \,\color{blue}{ \bar{\partial }\,C(x-y)\,A^b(y)\, \bar{\partial}\, C(y-x) \,A^b(x)}\nn\\
&& \hspace{2 cm}  +\,\color{red}{\del^+ C(x-y)\frac{\bar{\partial}}{\del^+} A^b(y)\,\del^+ C(y-x)\frac{\bar{\partial}}{\del^+} A^b(x)} \nn\\
&&\hspace{2 cm}  -\,2\,\color{orange}{\bar{\partial}\,C(x-y)\,A^b(y) \del^+ C(y-x)\frac{\bar{\partial}}{\del^+} A^b(x)}\,\nn\\
&& \hspace{2 cm} + \, \color{teal}{\frac{\partial \bar{\del}}{\del^+} C(x-y) \bar{A}^b(y)\, \del^+ C(y-x) A^c(x)}\nn\\
&&\hspace{2 cm}  -\,\color{purple}{ {\partial }\,C(x-y)\,\bar{A}^b(y)\,\del^+ C(y-x)\frac{\bar{\partial}}{\del^+} A^b(x)} \nn\\
&&\hspace{2 cm}  -\, \color{olive}{\bar{\partial }\,C(x-y)\,{A}^b(y)\,\del^+ C(y-x)\frac{{\partial}}{\del^+} \bar{A}^b(x)} \nn\\
 &&\hspace{2 cm}  +\,{\color{magenta} \del^+ C(x-y)\frac{{\partial}}{\del^+} \bar{A}^b(x)\,\del^+ C(y-x)\frac{\bar{\partial}}{\del^+} A^b(x)}{\biggl \}} + c.c.\;.\nn\\
\eea

\ndt Here $i,j$ run over the transverse variables $x,\bar{x}$.

\subsection{The fermion determinant}

\ndt Computing the fermion determinant is complicated by the presence of a four fermion interaction term in (\ref{Final L}). It was shown in~\cite{Casarin:2023xic} that the construction of Nicolai maps can also be extended to supersymmetric theories with four fermion interaction terms. In Appendix B, we explain why this term does not contribute to the fermion determinant at order $g^2$. The quadratic operator\footnote{Note that the zero modes of the operator $\del^+$ can be removed using the appropriate boundary conditions~\cite{Leibbrandt:1987qv}} $Q^{ac}$ relevant to order $g^2$ (again dropping terms which vanish after `trace-ing'), is 
\bea
Q^{ac}(x;A)&=&\frac{\Box}{\del^+}\delta^{ac}-2gf^{abc}\frac{1}{\del^+}(\partial\bar{A^b}+\bar{\partial}A^b)+2gf^{abc}\frac{\bar{\partial}}{\del^+}(A^b +2gf^{abc}{\bar{A}^b}\frac{\partial}{\del^+}\ ,
\eea
which may be written as 
\bea
Q^{ac}(x,y;A)&=&\frac{\Box}{\del^+}\Biggl(\delta^{ac}+2gf^{abc}\int dy\, \del^+ C(x-y)\frac{1}{\del^+}(\partial\bar{A}^b(y)+\bar{\partial}A^b(y))\nn\\
&&-2gf^{abc}\int dy \,\bar{\partial}C(x-y)A^b(y) -2gf^{abc}\int dy\,\del^+ C(x-y){\bar{A}^b(y)}\frac{\partial^{(y)}}{\del^+}\Biggr)\,.\nn\\
\eea
As $\det (Q)=\det(\Box/\del^+)\times \det(1+\textbf{Y})$, we use (\ref{Identity 1}) to compute the fermion determinant order by order in $g$ upto an overall constant $\det(\Box/\del^+)$. \\

\ndt The fermion determinant to order $g^2$ is
\bea
\label{fermi}
\hspace{-0.6 cm} \log \det(1+\textbf{Y})&=&2ng^2 \int dx\, dy\,\,{\biggl \{} \,\color{blue}{ \bar{\partial }\,C(x-y)\,A^b(y)\, \bar{\partial}\, C(y-x) \,A^b(x)}\nn\\
&& \hspace{2 cm}  +\,\color{red}{\del^+ C(x-y)\frac{\bar{\partial}}{\del^+} A^b(y)\,\del^+ C(y-x)\frac{\bar{\partial}}{\del^+} A^b(x)} \nn\\
&&\hspace{2 cm}  -\,2\,\color{orange}{\bar{\partial}\,C(x-y)\,A^b(y) \del^+ C(y-x)\frac{\bar{\partial}}{\del^+} A^b(x)}\,\nn\\
&& \hspace{2 cm} + \, \color{teal}{\frac{\partial \bar{\del}}{\del^+} C(x-y) \bar{A}^b(y)\, \del^+ C(y-x) A^c(x)}\nn\\
&&\hspace{2 cm}  -\,\color{purple}{ {\partial }\,C(x-y)\,\bar{A}^b(y)\,\del^+ C(y-x)\frac{\bar{\partial}}{\del^+} A^b(x)} \nn\\
&&\hspace{2 cm}  -\, \color{olive}{\bar{\partial }\,C(x-y)\,{A}^b(y)\,\del^+ C(y-x)\frac{{\partial}}{\del^+} \bar{A}^b(x)} \nn\\
 &&\hspace{2 cm}  +\,{\color{magenta} \del^+ C(x-y)\frac{{\partial}}{\del^+} \bar{A}^b(x)\,\del^+ C(y-x)\frac{\bar{\partial}}{\del^+} A^b(x)}{\biggl \}} + c.c.\;.
\eea
\ndt Thus, we find that the Jacobi determinant of the bosonic transformation (\ref{Jacobian}) exactly matches the fermion determinant (\ref{fermi}) upto $O(g^2)$.\\

\section{Extension of the light-cone map to all critical dimensions}
\ndt We now illustrate how this light-cone realization of the Nicolai map extends nicely to all critical dimensions (\ref{Tmap}). One perhaps obvious observation is that such a result cannot begin from a Lagrangian in a helicity basis, which is closely tied to four dimensions (and the little group $SO(2)$).

\subsection{The move away from a helicity basis}
 We start instead from the Lagrangian (\ref {L4}) in Appendix A. Note that the free bosonic and the free fermionic degrees of freedom match only in $d=3,4,6,10$. We show that when interactions are switched on, the fermion determinant matches the Jacobian only in the critical dimensions, hence confirming the existence of the supersymmetric Yang-Mills theories in these dimensions. \\
 
 \ndt We identify a field transformation for the physical fields $A_i$ so that we can write the pure Yang-Mills theory (first two lines of equation (\ref{L4})) as $\frac{1}{2} {A_i}^{\prime\,a}\Box A_i^{\prime\,a}$. In this section, all measures and delta functions will be $d$-dimensional (dimensions will get fixed using the determinant matching). Again, we introduce $\Box C(x-y)=-\delta^{(d)}(x-y)$ to write the map to order $g^2$
\bea
\label{Tmap}
\hspace{-0.5cm}A_i^{'\, a}(x)\!\!\!\!&=&A_i^a(x)+ gf^{abc} \int dy \, \left(\del^+ C(x-y) \frac{\pa_j}{\del^+} A_j^b(y)A_i^c(y) -\,\del_j C(x-y) A_j^b(y) A_i^c(y) \right)\nn\\
&&-\,\, \frac{g^2}{2} f^{abc}f^{bde}  \int dy \,\,\del^+ C(x-y) A_i^c(y) \frac{1}{\pa^{+ \,2}}\left(\pa^+ A_j^d(y) A_j^e(y)\right) \nn\\
&& +\,\, \frac{g^2}{2} f^{abc}f^{bde}  \int dy \,dz\,\, \biggl\{\, \del_j C(x-y)  A_k^c(y) \nn\\
&&\hspace{4cm}\times\,\,\left( \del_i C(y-z) A_k^d(z)A_j^e(z)+ \del_k C(y-z) A_j^d(z)A_i^e(z)\right) \nn\\
&& \hspace{4cm} -\,\,\del_i C(x-y) A_j^c(y) \,\pa^+ C(y-z) \frac{\pa_k}{\del^+} A_k^d(z) A_j^e(z)\nn\\
 &&\hspace{4cm}+\,\, \pa^+ C(x-y) \frac{\pa_j}{\del^+} A_j^c(y) \,\pa^+ C(y-z) \frac{\pa_k}{\del^+} A_k^d(z) A_i^e(z) \nn\\
&&\hspace{4cm}+ \,\,2 \, \del_i C(x-y)  A_j^c(y)  \del_k C(y-z) A_k^d(z)A_j^e(z)\nn\\
&& \hspace{4cm}-\,\, 2\,\pa^+ C(x-y) \frac{\pa_j}{\del^+} A_j^c(y)\, \del_k C(y-z) A_k^d(z)A_i^e(z) \nn\\
&&\hspace{4cm} +\,\, \pa^- C(x-y) A_k^c(y) {\del^+}C(y-z) A_k^d(z) A_i^e(z) \nn\\
&&\hspace{4cm} +\,\, \pa^+ C(x-y) A_k^c(y){\del^-}C(y-z) A_k^d(z) A_i^e(z) \nn\\
&&\hspace{4cm}-\,\, \del_j C(x-y)  A_k^c(y)  \del_j C(y-z) A_k^d(z)A_i^e(z)
\biggl\}\,. \nn\\
\eea

\subsection{Jacobian }
\ndt We calculate below the Jacobian of the transformation (\ref{Tmap}). 
\bea
\frac{\delta A_i^{'\,a}\,(x)}{\delta A_m^{\;p}\,(w)}&=&\delta_i^m \delta^{ap}\delta(x-w)+gf^{abc} \int dy\, \biggl\{ \del^+ C(x-y)  \delta(y-w) \biggl( \frac{\pa_j}{\del^+} \delta_j^m \delta^{bp}A_i^c(y)\nn\\
&& +\,\, \frac{\pa_j}{\del^+} A_j^b(y)  \delta_i^m \delta^{cp} \biggr) -\del_j C(x-y)  \delta(y-w) \left( \delta_j^m \delta^{bp} A_i^c(y) +  A_j^b(y)  \delta_i^m \delta^{cp} \right)  \biggl\} \nn\\
&&+ \,\,\frac{g^2}{2} f^{abc}f^{bde}  \int dy \,dz\, \Biggl\{ \del_j C(x-y)  A_k^c(y)\, \delta(z-w) \biggl\{\del_i C(y-z) \delta_k^m \delta^{dp} A_j^e(z)\nn\\
&& +\,\, \del_i C(y-z) A_k^d(z) \delta_j^m \delta^{ep} + \, \del_k C(y-z) \delta_j^m \delta^{dp} A_i^e(z)\nn\\
&&+\,\, \del_k C(y-z) A_j^d(z) \delta_i^m \delta^{ep}  \biggl\} -  \,\,\del_i C(x-y) A_j^c(y) \delta(z-w)\nn\\
&&\times\,\, \left( \,\pa^+ C(y-z) \frac{\pa_k}{\del^+} \delta_k^m \delta^{dp} A_j^e(z) +\, \pa^+ C(y-z) \frac{\pa_k}{\del^+} A_k^d(z) \delta_j^m \delta^{ep} \right) \nn\\
&& +\,\, \pa^+ C(x-y) \frac{\pa_j}{\del^+} A_j^c(y) \,\delta(z-w)\,\pa^+ C(y-z)\nn\\
&&\times\,\, \left(\, \frac{\pa_k}{\del^+} \delta_k^m \delta^{dp}  A_i^e(z) + \frac{\pa_k}{\del^+} A_k^d(z) \delta_i^m \delta^{ep} \,\right) + 2 \, \del_i C(x-y)  A_j^c(y) \,\delta(z-w)\nn\\
&&\times\,\, \left( \del_k C(y-z) \delta_k^m \delta^{dp} A_j^e(z) + \del_k C(y-z) A_k^d(z)  \delta_j^m \delta^{ep} \right) \nn\\
&&- \,\,2\,\pa^+ C(x-y) \frac{\pa_j}{\del^+} A_j^c(y)\, \delta(z-w)\nn\\
&&\times\,\, \left( \del_k C(y-z) \delta_k^m \delta^{dp} A_i^e(z) + \del_k C(y-z) A_k^d(z)\delta_i^m \delta^{ep} \right)\nn\\
&&+\,\,\pa^- C(x-y) A_k^c(y)\,  \delta(z-w)\nn\\
&&\times\,\,\left({\del^+} C(y-z) A_k^d(z) \delta_i^m \delta^{ep} +  {\del^+} C(y-z) A_i^e(z) \delta_k^m \delta^{dp} \right) \nn\\
&&+\, \,\pa^+ C(x-y) A_k^c(y)\, \delta(z-w)\nn\\
&&\times\,\,\left( \del^- C(y-z) A_k^d(z) \delta_i^m \delta^{ep} +  {\del^-} C(y-z) A_i^e(z) \delta_k^m \delta^{dp} \right)\nn\\
&& -\,\, \del_j C(x-y)  A_k^c(y) \, \delta(z-w)\nn\\
&&\times\,\,\left( \del_j C(y-z)  \delta_k^m \delta^{dp} A_i^e(z) +  \del_j C(y-z)   A_k^d(z) \delta_i^m \delta^{ep} \right) \biggr\}\,,\nn\\
\eea
where we have only written the non-trivial terms relevant till order $g^2$. 
We take the trace by setting $a=m$, $w=x$ and integrating over $x$. We also use the $SU(n)$ identity $f^{abc}f^{abd} = n\delta^{cd}$ to obtain the Jacobi determinant upto $O(g^2)$ using (\ref{Identity 1})
\bea
\label{Jacobian1}
\hspace{-1.2 cm} \log \det\left(\frac{\delta A_i'^a(x)}{\delta A_j^m(w) }\right)&=& ng^2 \int dx\, dy\,\, (d-2){\biggl \{} {\color{blue} \,{ {\partial_i }\,C(x-y)\,A_i^b(y)\, {\partial_j}\, C(y-x) \,A_j^b(x)}}\nn\\
&& \hspace{2 cm}  {\color{red}+\, \,{\partial^+C(x-y)  \frac{\pa_j}{\del^+} A_j^b(y)\,\partial^+C(y-x) \frac{\pa_i}{\del^+}A_i^b(x)} }\nn\\
&&\hspace{2 cm}  {\color{orange}-\,2\,\,{\partial_i }\,C(x-y)\,A_i^b(y) \partial^+C(y-x) \frac{\pa_j}{\del^+}A_j^b(x)}\,\nn\\
&& \hspace{2 cm} {\color{teal} - \frac{1}{2} \,{\partial_i }\,C(x-y)\,A_j^b(y)\, {\partial_i}\, C(y-x) \,A_j^b(x)}\nn\\
&&\hspace{2 cm}  {\color{purple}+\,  \frac{1}{2}\,\frac{\del_i^2}{\del^+} C(x-y) A_j^b(y)\, \del^+ C(y-x) A_j^b(y) } \biggl\}\,, \nn\\
\eea
\ndt where we have used the relation $2\,\del^-C(x-y) = \frac{\del_i^2}{\del^+} C(x-y)+ \frac{1}{\del^+} \delta(x-y)$. \\

\subsection{Fermion Determinant}
As noted in section 2, the presence of the four fermion interaction term in (\ref{L4}) makes the computation of the fermion determinant slightly involved. But the contribution from such a term to the fermion determinant is trivial at order $g^2$ as shown in Appendix B. \\

\ndt We simplify the quadratic operator in (\ref{L4}) by expanding the covariant derivatives and using the constraint equation (\ref{constr}). We get
\bea
&&\Delta=\det\biggl\{\frac{1}{2}\,\frac{\Box}{\partial^+}\delta^{ac}- \frac{1}{2}gf^{abc}\, \gamma^i\gamma^j \frac{\partial_i}{\partial^+}\,(A_j^b\,-\, \frac{1}{2}gf^{abc}\,\gamma^i\gamma^j A_i^b \frac{\partial_j}{\partial^+ }\, -  gf^{abc}\,\frac{\partial_i}{\partial^+}A_i^b \nn\\
&& \hspace{2 cm} - \, g^2f^{abc}f^{bde}\,\,\frac{1}{\partial^{+\,2}}\left(A_i^d\partial^+A_i^e\right) -\frac{1}{2}g^2f^{ade}f^{bcd}\gamma^i\gamma^j A_i^e \frac{1}{\partial^+}A_j^b\biggl\}\,.
\eea
\ndt The non-trivial part of the quadratic operator relevant to order $g^2$ is
\bea
\label{FD1}
&&\Delta =\det(\frac{1}{2}\,\frac{\Box}{\partial^+})\cdot\det\biggl\{\delta^{ac}+gf^{abc}\gamma^i\gamma^j \int \!dy \, \del^+C(x-y)A_i^b(y)\,\frac{\del_j}{\del^+}\nn\\
&& + gf^{abc}\gamma^i\gamma^j \int \!dy\, \del_iC(x-y)  A_j^b(y)  +2\,gf^{abc}  \,\int \!dy\,\del^+ C(x-y)\frac{\del_i}{\del^+} A_i^b(y) \biggl\}\,.\nn\\
\eea
\ndt We now compute the fermion determinant  pertubatively using the (\ref{Identity 1}). We use the trace relation satisfied by the gamma matrices given in appendix A (\ref{Tr}) to obtain
\bea
\label{FD}
\hspace{-0.6 cm} \log \det(1+{\bf Y})&=& ng^2 \int dx\, dy\,\, \frac{r}{4}{\biggl \{} {\color{blue} \,2\,{ {\partial_i }\,C(x-y)\,A_i^b(y)\, {\partial_j}\, C(y-x) \,A_j^b(x)}}\nn\\
&& \hspace{2 cm}  {\color{red}+\,2 \,{\partial^+C(x-y)  \frac{\pa_j}{\del^+} A_j^b(y)\,\partial^+C(y-x) \frac{\pa_i}{\del^+}A_i^b(x)} }\nn\\
&&\hspace{2 cm}  {\color{orange}-\,4\,\,{\partial_i }\,C(x-y)\,A_i^b(y) \partial^+C(y-x) \frac{\pa_j}{\del^+}A_j^b(x)}\,\nn\\
&& \hspace{2 cm} {\color{teal} -\,\,{\partial_i }\,C(x-y)\,A_j^b(y)\, {\partial_i}\, C(y-x) \,A_j^b(x)}\nn\\
&&\hspace{2 cm}  {\color{purple}+\,\, \frac{\del_i^2}{\del^+} C(x-y) A_j^b(y)\, \del^+ C(y-x) A_j^b(y) } \biggl\}\,,\nn\\
\eea
where $r={\rm Tr}\,{\bf 1}\;  ({\bf 1}$ is the identity matrix) and it counts the number of off-shell fermionic degrees of freedom.
\subsection{Existence of map in critical dimensions}
\ndt We see now that the Jacobian determinant (\ref{Jacobian1}) matches against the fermion determinant (\ref{FD}) if and only if
\bea
&&{\color{blue}  \frac{r}{2} = d-2 }\,,  \nn\\ 
&& {\color{red} \frac{r}{2} = d-2 }  \,,\nn\\
&&{\color{orange}\! \!\!-r = - 2(d-2) }\,, \nn\\
&&{\color{teal}\! \!\!\!-\frac{r}{4} = - \frac{d-2}{2} }\,, \nn\\
&& {\color{purple} \!\!\!\! +\frac{r}{4} = \frac{d-2}{2}  }\,, \nn
\eea
\ndt all implying that $r=2(d-2)$ which happens for $d=3,4,6$ and $10$ \cite{Ananth:2020gkt,Ananth:2020lup,Malcha:2021ess}. We find that the matching of determinants depends on the dimension of our field theory, which imposes a constraint on the allowed values of space-time dimensions. We recover the old result using this approach that the pure supersymmetric Yang-Mills theories can only exist in $d=3,4,6,10$ dimensions. This result was first obtained in \cite{Brink:1976bc} using the closure of supersymmetry transformations and requiring the use of specific Fierz identity. Hence supersymmetric Yang-Mills theories can be formulated in the light-cone gauge in all critical dimensions without using anti-commuting variables. \\

\ndt Note that the constraints on the fermion fields are dimension dependent. For example in $d=3,4$, the fermion fields are Majorana spinors, in $d=6$  Weyl spinors, while in $d=10$, they are Majorana-Weyl spinors.\\

\subsection{Comparison between maps}
\ndt We briefly explain the connection between our light-cone maps and the map constructed in ref.~\cite{Malcha:2021ess}. We note that we can recover our map for the physical fields $A_i$ (\ref{Tmap}) (to order $g$) from the one derived in~\cite{Malcha:2021ess} in general gauges ($n^{\mu}A_{\mu}=0$). Note that in ref.~\cite{Malcha:2021ess}, the map is written covariantly. In our case, we have lost manifest covariance due to the elimination of unphysical degrees of freedom. Hence, we can only match the transverse part of the map in \cite{Malcha:2021ess} by employing the following conditions in equation (4.1) of ref.~\cite{Malcha:2021ess}: $n^{\mu}$ is chosen to be null. In the light cone coordinates, this means setting the components $n^+=n^i=0$ and $n^-=1$. The ghost propagator (in the light-cone gauge) is  $\frac{1}{\del^+}$. Using the above conditions and the constraint equation ((\ref{constr}) from the appendix) without the fermion term, the map in ref.\cite{Malcha:2021ess} matches with our map (\ref{Tmap}) to the cubic order. \\

\ndt Recently, in \cite{Lechtenfeld:2024uhi}, a four-dimensional and ten-dimensional Nicolai map was constructed using the $\mathcal{R}$ operator in the light-cone gauge. The four-dimensional map eq 2.17 in \cite{Lechtenfeld:2024uhi} is different than (\ref{Tnmap}), and they cannot be compared due to the different structures. This distinction can be understood at the level of Lagrangian which we discuss in detail in the next section. However, we can obtain their four-dimensional map by taking the equation (\ref{Tmap}) in four dimensions and transforming only one of the helicity variable ($A$). In ten dimensions, both the maps (\ref{Tmap}) and eq 3.16 in \cite{Lechtenfeld:2024uhi}  are the same till order $g$, and at order $g^2$, the maps differ.  \\

\ndt The free action condition is satisfied differently due to the freedom in writing the Lagrangian using partial integrations. Similarly, the determinant (Jacobian) depends on the form of the map, and many cancellations between the terms happen at the Jacobian level. \\

\section{On the issue of uniqueness}
\ndt The map obtained in the preceding section (\ref{Tmap}) can be written in $d=4$ in terms of the helicity variables and fields. This is distinct from the map (\ref{Tnmap}) obtained in section 2. One can understand this distinction at the level of Lagrangian. The cubic terms (purely involving the gauge fields) in (\ref{Final L}) and (\ref{L4}) are related by partial integrations in $d=4$. However, the maps (\ref{Tnmap}) and (\ref{Tmap}) to the cubic order cannot be connected by any partial integrations. Structurally the map (\ref{Tnmap}) is of the form $\del\,C\,A\,A$ while the map (\ref{Tmap}) is of the form $\del\,C\,(\,A\,A\,+\,A\,\bar{A})$ to the cubic order ($\del$ collectively represents the space-time derivatives). Hence, these two maps are distinct. \\

\ndt We can, in principle, write distinct maps by writing the Lagrangian in different ways by performing partial integrations. The non-trivial check required for each map is determinant matching. In our case, we find that there are two maps at order $g^2$ that satisfy all the conditions of the Main theorem (listed in section 2). The determinant matching condition is about the equality of the derivative (Jacobian) of the map with the fermion determinant and not about the map itself, hence the non-uniqueness. Moving to higher orders may fix this uniqueness issue. It is useful to point out that when one works to any finite order in a perturbation theory, one can always find the simplest map relevant to that order, which will simplify the computation of correlation functions (scattering amplitudes). \\

\ndt   We find that the two maps can be related  at the level of the Jacobian\footnote{The two six dimensional maps in \cite{Ananth:2020lup,Ananth:2020jdr} can {\it {also}} be related to each other at the level of the Jacobi determinant.  To do this, we simply need to add terms proportional to $(d-6)$ to the Jacobian of the six dimension map (obtained by trial and error) in \cite{Ananth:2020jdr}.}. We can take the expression (\ref{Jacobian1}) (the Jacobian of the map (\ref{Tmap})) and write it in the four dimensional helicity variables. We find that the expression (\ref{Jacobian}) (the Jacobian of the map (\ref{Tnmap})) is recovered after cancellation of some terms. This implies that the four dimensional map (\ref{Tnmap}) is `simple' as it does not produce any extra terms to be cancelled in the Jacobian.  Thus, to any given order one can find different maps which satisfy all the conditions of the main theorem.\\

\subsection*{Towards $\mathcal{N}=1$ light-cone superspace in $d=4$}

\ndt In the light-cone gauge, we found that the Lagrangian (\ref{Final L}) contains a four fermion interaction term. This might complicate the construction of Nicolai map at higher orders in the coupling. An alternative way out is to study this approach in the language of superspace.  In the following section, we find a superspace map and comment on its properties.

\section{The map in superspace}

\ndt In the light-cone superspace, the Lagrangian (\ref{Final L}) takes a very simple form. The main ingredient in superspace is the superfield given by
\bea
\label{sf}
\phi(y,\theta)=iA(y)+\theta \bar{\chi}(y)\,.
\eea

\ndt where $y=(x^+,x^--\frac{i}{\sqrt{2}}\theta\bar{\theta},x,\bar{x})$ and $\theta,\bar{\theta}$ are the anti-commuting variables. We also have the covariant derivatives $\mathrm{d}\,,\bar{\mathrm{d}}$ given by
\bea
\mathrm{d}=-\frac{\partial}{\partial \bar{\theta}}-\frac{i}{\sqrt{2}}\theta \partial^+\,, \hspace{1cm} \bar{\mathrm{d}}=\frac{\partial}{\partial \theta}+\frac{i}{\sqrt{2}}\bar{\theta} \partial^+\,,
\eea 

\ndt which satisfy $\{\mathrm{d},\bar{\mathrm{d}}\,\}=-i\sqrt{2}\partial^+$. The superfield and its conjugate satisfy $\mathrm{d}\phi=0$ and  $\bar{\mathrm{d}}\bar{\phi}=0$. The action is then
\bea
\label{superl}
\mathcal{S}=\int d^4x \,d^2\theta\, \bigg\{-\frac{i}{\sqrt{2}}\bar{\phi}^a\frac{\Box}{\partial^+}\phi^a +\sqrt{2}gf^{abc}\left(\frac{\bar{\partial}}{\partial^+}\phi^a\phi^b\bar{\phi}^c+\frac{\partial}{\partial^+}\bar{\phi}^a\bar{\phi}^b\phi^c\right)\nn\\
\hspace{0.3cm}-g^2f^{abc}f^{ade}\frac{1}{\partial^+}(\phi^b\, \mathrm{d}\bar{\phi}^c)\frac{1}{\partial^+}(\bar{\phi}^d\,\bar{\mathrm{d}}\phi^e)\bigg\}\,. \nn\\
\eea

\ndt In the following, we adopt the notation : $\textbf{x}=(y,\theta)$ and $\widetilde{\textbf{x}}=(x,\theta,\bar{\theta)}$. The former will be referred to as `chiral coordinates' while the latter as `super coordinates'. The chiral superfield in the Lagrangian thus depends on the chiral coordinates as $\phi^a(\textbf{x})$. The integration measure is written as $d\widetilde{\textbf{x}}$ . We find a transformation to order $g^2$ such that it maps the full $\mathcal{N}=1$ Lagrangian to a free Lagrangian in superspace.  The Jacobi determinant of the transformation is one and it preserves the chirality of the superfield. In order to write such a transformation, we use
\bea
\label{Gf}
 \Box G(\widetilde{\textbf{x}}-\widetilde{\textbf{y}})=\frac{\mathrm{d\bar{d}}}{-i\sqrt{2}\del^+}\delta^{(6)}(\widetilde{\textbf{x}}-\widetilde{\textbf{y}})\,.
 \eea
 \ndt Note that $G(\widetilde{\textbf{x}}-\widetilde{\textbf{y}})$ is chiral w.r.t $\widetilde{\textbf{x}}$ while it is anti-chiral w.r.t. $\widetilde{\textbf{y}}$. \\
 
\ndt We find the following transformation
\bea
\label{smap}
\phi^{'\,a}(\textbf{x})=&&\phi^a(\textbf{x})+2igf^{abc}\int d\widetilde{\textbf{y}} \,\partial^{+}G(\widetilde{\textbf{x}}-\widetilde{\textbf{y}})\frac{\bar{\partial}}{\partial^+}\phi^{b}(\textbf{y})\phi^c(\textbf{y})\nn\\
&&+\frac{i}{\sqrt{2}} g^2 f^{abc}f^{bde}\int d\widetilde{\textbf{y}} \,\partial^{+}G(\widetilde{\textbf{x}}-\widetilde{\textbf{y}})\,\,\bar{\mathrm{d}}\phi^c(\textbf{y)}\frac{1}{\partial^{+\,2}}[\phi^d(\textbf{y})\, \mathrm{d}\bar{\phi^e}(\bar{\textbf{y}})]\nn\\
&&\,+\,2g^2f^{abc}f^{bde}\int d\widetilde{\textbf{y}} \,d\widetilde{\textbf{z}}\, \partial^{+}G(\widetilde{\textbf{x}}-\widetilde{\textbf{y}})\frac{\partial}{\partial^+}\bar{\phi^c}(\bar{\textbf{y}})\,\partial^{+}G(\widetilde{\textbf{y}}-\widetilde{\textbf{z}})\frac{\bar{\partial}}{\partial^+}\phi^{d}(\textbf{z})\phi^e(\textbf{z})\,,\nn\\
\eea

\ndt where $\textbf{x}=(y,\theta)$, $\textbf{y}=(y^{'},\theta^{'})$, $\textbf{z}=(y^{''},\theta^{''})$ are the chiral coordinates and $\widetilde{\textbf{y}}=(x^{'},\theta^{'},\bar{\theta}^{'})\,,\,\widetilde{\textbf{z}}=(x^{''},\theta^{''},\bar{\theta}^{\,''})$ are the super coordinates. The map for $\bar{\phi}^{\,'\,a}$ is obtained by complex conjugation.\\

 \ndt It is easy to check that we can now write the Lagrangian as $-\frac{i}{\sqrt{2}}\,\bar{\phi}^{'\,a}\frac{\Box}{\partial^+}\phi^{'\,a}$. We compute the Jacobian of the transformation using the fact that
 \bea
\frac{ \delta \phi^a(\textbf{x})}{\delta \phi^b(\textbf{y})}=\delta_b^a\,\mathrm \,\delta(\textbf{x}-\textbf{y})\,,\,\hspace{1cm}\frac{ \delta \phi^a(\textbf{x})}{\delta \bar{\phi}^b(\bar{\textbf{y}})}=0\,.
 \eea
 
\ndt The delta functions of the chiral coordinates are $5$-dimensional while those of the super coordinates are $6$-dimensional.  Note that the above functional variation is chiral.  We can write the delta function in terms of $\widetilde{\textbf{x}}$ and $\widetilde{\textbf{y}}$ as
\bea
\label{nid}
\delta(\textbf{x}-\textbf{y})=\text{d}\delta(\widetilde{\textbf{x}}-\widetilde{\textbf{y}})\,.
\eea

\ndt The non-trivial contributions to the Jacobian relevant at order $g^2$ are
\bea
\frac{ \delta \phi^{'\,a}(\textbf{x})}{\delta \phi^{\,m}(\textbf{w})}&=&\delta_m^a\,\delta(\textbf{x}-\textbf{w})\nn\\
&&+\,2igf^{abc}\int d\widetilde{\textbf{y}} \,\partial^{+}G(\widetilde{\textbf{x}}-\widetilde{\textbf{y}}){\biggl \{}\delta_m^b \frac{\bar{\partial}}{\partial^+}(\delta(\textbf{y}-\textbf{w}))\phi^c(\textbf{y})\nn\\
&&\hspace{4cm}+\frac{\bar{\partial}}{\partial^+}\phi^b(\textbf{y})\,\delta_m^c \,\delta(\textbf{y}-\textbf{w}){\biggl \}}\nn\\
&&\,+\,2g^2f^{abc}f^{bde}\int d\widetilde{\textbf{y}} \,d\widetilde{\textbf{z}}\, \partial^{+}G(\widetilde{\textbf{x}}-\widetilde{\textbf{y}})\frac{\partial}{\partial^+}\bar{\phi}^c(\textbf{y})\,\partial^{+}G(\widetilde{\textbf{y}}-\widetilde{\textbf{z}})\nn\\
&&\hspace{2cm}\times\,\,{\biggl \{ }\delta_m^d\frac{\bar{\partial}}{\partial^+}\,(\delta(\textbf{z}-\textbf{w}))\phi^e(\textbf{z})+\frac{\bar{\partial}}{\partial^+}\phi^{d}(\textbf{z})\,\delta_m^e\,\delta(\textbf{z}-\textbf{w}){\biggr \}}\,.\nn\\
\eea

\ndt Using (\ref{Identity 1}) along with (\ref{nid}) and the fact that $\mathrm{d} G(\widetilde{\textbf{x}}-\widetilde{\textbf{y}})=0$ yields
 \bea
\det\left(\frac{ \delta \Phi_i^{'\,a}(\widetilde{\textbf{x}})}{\delta \Phi_j^{\,m}(\widetilde{\textbf{w}})}\right)=1+ \mathcal{O}(g^3)\,,
 \eea
 \ndt where $i,j=1,2$ and $\Phi_1\equiv\phi,\;\Phi_2\equiv\bar{\phi}$. Thus we find that by using the chirality constraint in superspace, the trace vanishes till order $g^2$. We expect this to hold at higher orders in the coupling as well. 
 
\subsection{Component maps}
\ndt Here, in this subsection, we present the component maps obtained from the  superspace map (\ref{smap}). We start with the transformation (\ref{smap}), plug the form of the superfield (\ref{sf}), and extract the maps for each component field.  \\

\ndt To obtain the map in terms of component fields, we need the form of Green's function $G(\widetilde{\textbf{x}}-\widetilde{\textbf{y}})$. Using (\ref{Gf}), we get
\bea
\label{gff}
G(\widetilde{\textbf{x}}-\widetilde{\textbf{y}}) = \frac{d \bar{d}}{i\sqrt{2} \del^+} \,[ C(x-x')\, \delta^2(\theta -\theta')\,]
\eea
where $C(x-x')$ is the scalar Green's function as defined in section (\ref{Trans}), and the two-dimensional delta function involves both $\theta$ and $\bar{\theta}$ variables. \\

\ndt For the sake of simplicity, we start with the superspace transformation (\ref{smap}) only till order $g$ in the coupling
\bea
\phi^{'\,a}(\textbf{x})=\phi^a(\textbf{x})+2igf^{abc}\int d\widetilde{\textbf{y}} \,\partial^{+}G(\widetilde{\textbf{x}}-\widetilde{\textbf{y}})\frac{\bar{\partial}}{\partial^+}\phi^{b}(\textbf{y})\phi^c(\textbf{y})\nn
\eea
Plugging the form of the superfield (\ref{sf}) and using the Green's function definition (\ref{gff}), we get the following component maps till order $g$
\bea
&&A^{'\,a}(g;A,\bar{A})=A^a(x)+2gf^{abc} \int dy \,\del^+C(x-y) \frac{\bar{\partial}}{\del^+}\,A^b(y)A^c(y)\nn\\
&&{\chi}^{'\,a}(g;\bar{A},{\chi})= {\chi}^a(x) + 2gf^{abc} \int dy\, \del^+C(x-y)\, \left( \frac{{\partial}}{\del^+}\,\bar{A}^b(y) {\chi}^c(y) + \frac{{\partial}}{\del^+}\,{\chi}^b(y) \bar{A}^c(y) \right) \nn\\
\eea
Here, we find field transformations for both gauge and fermion fields that map the interacting super Yang-Mills theory to a free theory. Note that this is an artifact of working with the light-cone superspace; in this formalism, all component fields are on an equal footing. The gauge field map obtained here matches with the result (\ref{Tnmap}) derived in section (\ref{Trans}). The transformation for the fermion field is a surprising entry here.\\

\ndt One can extend this construction till order $g^2$ and find that the Jacobian of both the field transformations will cancel order by order in the coupling. Therefore, in this procedure, the interacting supersymmetric theories can be mapped to a free supersymmetric theory
in a different way as compared to the Nicolai map formalism.\\

\ndt Hence, the superspace map offers an alternative way to compute correlation functions using a free theory in superspace. Finally, the goal would be to find such a map for the maximally supersymmetric Yang-Mills theory and see what it can teach us about the mathematical properties of the theory.

\section{ Discussion on Nicolai map and quadratic forms}
\ndt The light-cone Hamiltonians for  the pure and the maximally supersymmetric theories in $d=4$ can be expressed as quadratic forms \cite{Ananth:2005,Mali:2015,Ananth:2020mws}. Here, we focus on the quadratic form structure in the pure Yang-Mills theory. The light-cone Hamiltonian for this theory may be written as
\bea
\mathcal{H}=2\int d^3x \, \mathcal{D}\bar{A}^a\bar{\mathcal{D}}A^a\,,
\eea  
\ndt where
\bea
\label{qf}
\bar{\mathcal{D}}A^a=\bar{\partial}A^a-gf^{abc}\frac{1}{\parm}(\bar{A}^b\parm A^c)\,,
\eea
\ndt and $\mathcal{D}\bar{A}^a$ is obtained by complex conjugation. \\

\ndt We now propose an alternative way to express this Hamiltonian. We introduce new variables $A'^{a}\,,\bar{A}'^{a}$ given by
\bea
\label{nmap}
A'^{a}(x)=A^a(x)-2gf^{abc}\int d^2y\,\partial C^T(x-y)\frac{1}{\parm}[\bar{A}^b(x')\parm A^c(x')]\,,
\eea
\ndt where $x'=(x^+,x^-,y,\bar{y})$ and $2\,\partial\bar{\partial}C^T(x-y)=-\delta^2(x-y)$. In these new variables, the Hamiltonian takes the form
\bea
\mathcal{H}=-2\int d^3x \,\bar{A'}^a \partial \bar{\partial} A'^a\,.
\eea

\ndt Thus, (\ref{nmap}) is a bosonic transformation that maps the Yang-Mills Hamiltonian to a free Hamiltonian. It is therefore of interest to explore whether this transformation (\ref{nmap}) has a connection with the Nicolai map (\ref{Tnmap}). We believe that the expression (\ref{nmap}) represents a good place to begin an investigation of possible links between the Nicolai map and quadratic form.

\section*{Acknowledgments}
We thank Sudarshan Ananth for many helpful discussions. SP and NB acknowledge support from the Prime Minister’s Research Fellowship (PMRF) and CSIR-NET fellowship respectively. \\

\newpage

\appendix
\section*{Appendix}
\vskip 0.2 cm
\section{Notations and conventions}

We start with a gauge theory involving bosonic ($A_{\mu}$) and fermionic ($\psi$) degrees of freedom in the adjoint representation. We work here in the light-cone gauge and do not explicitly specify the dimension.   The allowed dimensions for the supersymmetric Yang-Mills theories to exist will get fixed using the Nicolai map (particularly the matching of fermion and the Jacobi determinants). These dimensions are referred as the `critical dimensions' \cite{Ananth:2020lup}. \\


\ndt The light cone coordinates are given by
\bea
x^{\pm}=\frac{(x^0\,\pm \,x^{d-1})}{\sqrt{2}}  \,.
\eea

\ndt The transverse coordinates are given by $x_i$ where $i=1,...d-2$. The derivatives with respect to the light-cone coordinates are denoted by $\del_{\pm}$ ($-\del^{\mp}$) while those with respect to the transverse coordinates are denoted by $\del_i$.\\

\ndt Gamma matrices satisfy $\left\{\gamma^{\m}\, ,\,\gamma^{\nu}\right\}=-2\eta^{\mu \nu}$ where $\eta_{\mu\nu}$ is the light-cone metric. The $\gamma^{\pm}$ are defined as
\bea
\gamma^{\pm}=\frac{1}{\sqrt{2}}\left(\gamma^0\pm\gamma^{d-1}\right)\,. 
\eea
\ndt They satisfy
\bea
\gamma^{\pm \, 2}=0\,,\hspace{1 cm} \gamma^{+ ^{\,\dagger}}=\gamma^{-} \,, \hspace{1 cm} \left\{\gamma^{\pm}\,,\, \gamma^{i}\right\}=0\,,\hspace{1 cm}  \left\{\gamma^{+},\gamma^{-}\right\}=2\,.
\eea
\ndt The gamma matrices satisfy the following trace identities
\bea
\label{Tr}
{\rm Tr} \left( \gamma^{\m}\,\gamma^{\nu}\right)&=& - r \,\eta^{\mu \nu}\nn\\
{\rm Tr} \left( \gamma^{\m}\,\gamma^{\nu}\,\gamma^{\rho}\,\gamma^{\sigma}\right)&=&-r(\eta^{\mu\nu}\,\eta^{\rho\sigma}-\eta^{\mu\rho}\,\eta^{\nu\sigma}+\eta^{\mu\sigma}\,\eta^{\rho\nu})
\eea
where $r=2^\frac{[D]}{2} $ and it counts the number of off-shell fermionic degrees of freedom.\\

\ndt We introduce two hermitian projection operators
\bea
P_+=\frac{1}{2} \gamma^-\gamma^+ \;,\;\;\;\;\; P_-=\frac{1}{2} \gamma^+\gamma^- \;\;\; \text{which satisfy}\; \;P_{\pm}^{\,2}=P_{\pm} \;\;,\;\; P_+P_-=P_-P_+=0.
\eea

\vskip 2mm
\ndt We start with the action
\bea
\label{L2}
S= \int d^d x \;\left(-\frac{1}{4}F_{\mu\nu}^{\,a}F^{\mu\nu\,a}+ \frac{i}{2}\bar{\psi}^{a}\,\gamma^{\mu}\left(D_{\mu} \psi\right)^a \right)\,,
\eea
where $F_{\mu\nu}^{\,a}=\partial_{\mu}A_{\nu}^{\,a}-\partial_{\nu}A_{\mu}^{\,a}+gf^{abc}A_{\mu}^{\,b}A_{\nu}^{\,c}$ , $\bar{\psi}=\psi^{\dagger}\gamma^0$ and $D_\mu=\partial_\m\delta^{ac}+gf^{abc}A_\mu^b$ . The $f^{abc}$ are the structure constants of the gauge group $SU(n)$. \\

\ndt Here, we will not explicitly distinguish between Majorana, Weyl, and Majorana-Weyl spinors to keep notations simple. Note that this is justified because our calculations require only basic Clifford algebra and the trace relation (\ref{Tr}). \\

\ndt The equation of motion corresponding to the gauge field is
\bea
\label{eom}
D_{\mu} F^{\mu\nu\,a}-\frac{i}{2}gf^{abc}\bar{\psi}^{b}\,\gamma^{\nu}\psi^c=0\,.
\eea

\ndt  We make the gauge choice $A_-^{a}=0$ which renders $\nu=+$ in (\ref{eom}) as a constraint equation and we get
\bea
\label{constr}
A_+^a=-\frac{1}{\del^+}(\partial_i A^{i\,a})-gf^{abc}\frac{1}{\del^{+\,2}}(A_i^b\del^+ A^{i\,c})-\frac{i}{2}gf^{abc}\frac{1}{\del^+}(\bar{\psi}_+^a\gamma^+\psi_+^a)\,,
\eea

\ndt where the operation $\frac{1}{\del^+}$ is defined as

\bea
\frac{1}{\del^+}f(x^-)=-\int dy^-\,\, \theta(x^- - y^-)\,f(y^-)\,,
\eea
\ndt where $\theta(x^- - y^-)$ is the step function.\\

\ndt The equation of motion for the fermion field is given by $\gamma^\m D_\m^{ac} \psi^c=0.$  The fermion field $\psi$ can be decomposed into $\psi_{\pm}$ using the projection operators
\bea 
\psi_{\pm}= P_{\pm} \psi\;\;\;,\;\;\; \bar{\psi}_{\pm}= \bar{\psi}P_{\mp} \;\;\text{and}\;\; \psi =\psi_+\,+\psi_- \;\;,\;\;   \bar{\psi} =\bar{\psi}_+\,+\bar{\psi}_-\,.
\eea
Acting $P_+$ and $ P_-$ on the equation of motion, we get the following two equations
\bea
\label{fconstr}
&&D_-^{ac}\,\psi_-^c=-\frac{1}{2}\gamma^+\gamma^i \,D_i^{ac}\,\psi_+^{c}     \hspace{2 cm}   i=1,.....d-2  \,, \\
\label{Deqtn2}
&&D_+^{ac}\,\psi_+^{c}=-\frac{1}{2}\gamma^-\gamma^i \,D_i^{ac}\,\psi_-^{c}    \hspace{2 cm}   i=1,.....d-2\,.
\eea
Since  (\ref{fconstr}) is a constraint so we solve for $\psi_-^a$ and we obtain
\bea
\label{Deqtn3}
\psi_-^a=\frac{1}{2}\gamma^+\gamma^i\frac{1}{\partial^+} D_i^{ac} \psi_+^{c}\,.
\eea

\ndt Expanding the fermion term in terms of $\psi_{\pm}$ (and its conjugate) and substituting the constraint equations (\ref{constr}),(\ref{fconstr}) in (\ref{L2}), we obtain the Lagrangian in the light-cone gauge
\bea
\label{L4}
\mathcal{L}&=&\frac{1}{2}\,A_i^a \Box A_i^a - g f^{abc}\,\left(\frac{\pa_i }{\pa^+}A_i^a \,\pa^+ A_j^b A_j^c +\pa_i A_j^a A_i^b A_j^c \right) \nn\\
&& -g^2 f^{abc} f^{ade} \left(\; \frac{1}{4} A_i^b A_j^c A_i^d A_j^e + \frac{1}{2}\frac{1}{\partial^+}\left(\partial^+A_i^b A_i^c\right)\frac{1}{\partial^+}\left(\partial^+A_j^d A_j^e\right)\; \right) \nn\\
&&+ \frac{i}{2}\,\bar{\psi}_+^a\, \gamma^+\left(D_+^{\,ad}-\frac{1}{2}\,\gamma^i D_i^{\,ac}\frac{1}{\partial^+}\gamma^j D_j^{\,cd}\right)\psi_+^d \nn\\
&&-\frac{1}{8} \,g^2 f^{abc} f^{ade} \frac{1}{\partial^+}\,\left(\bar{\psi}_+^b \gamma^+ \psi_+^c \right)\,\frac{1}{\partial^+}\,\left(\bar{\psi}_+^d \gamma^+ \psi_+^e \right)\,.\nn\\
\eea

\ndt The last term of (\ref{L4}) is a four fermion interaction term which was absent in the original action. This is a feature which distinguishes the light-cone formulation of the theory from other gauges.\\ 

\ndt We now restrict to $d=4$ and  introduce the transverse coordinates 
\bea
x=\frac{(x^1+ix^2)}{\sqrt{2}}\,,\hspace{1cm} \bar{x}=x^*\,,
\eea
and their derivatives $\bar{\del}\,,\,\del$ respectively. We also introduce the helicity field
\bea
 A^a=\frac{A^a_1+iA^a_2}{\sqrt{2}}\,,
 \eea
 \ndt and its conjugate $\bar{A}^a$. \\

\ndt The $4 \times 4$ gamma matrices are 
\bea
\gamma^0=\begin{pmatrix}
0&&1\\
1&&0\\
\end{pmatrix}\;,\;\;\;\;
\gamma^i=\begin{pmatrix}
0&&\sigma^i\\
-\sigma^i&&0\\
\end{pmatrix}\, \;,
\eea
where $\sigma^i$ are the standard Pauli matrices. \\

\ndt The fermion fields in $d=4$ satisfy the Majorana condition $\psi = C \bar{\psi}^T$ where the charge conjugation matrix $C$ is 
\bea
C=\begin{pmatrix}
i\sigma^2&&0\\
0&&-i\sigma^2\\
\end{pmatrix}\,.
\eea

\ndt The fermion field $\psi^a(x)$ takes the form 

\bea
\psi=\begin{pmatrix}
\,\,\frac{\bar{\del}}{\del^+}\bar{\chi}\,\,\\[6pt]
-\bar{\chi}\\
\chi \\[6pt]
\frac{\del}{\del^+}\chi \\
\end{pmatrix}\,.
\eea

\ndt The Lagrangian (\ref{L4}) can now be simplified and written purely in terms of the physical fields ($A^a,\bar{A}^a,\chi^a,\bar{\chi}^a$) as

\bea
\label{Final LL}
\hspace{1 cm}\mathcal{L}&=&\bar{A}^a \Box A^a - 2g f^{abc} \left(\frac{\bar{\partial}}{\partial^+}A^a\partial^+\bar{A}^b A^c+\frac{\partial}{\partial^+}\bar{A}^a\partial^+A^b\bar{A}^c\right) \nn\\
&& \hspace{1 cm} -2g^2 f^{abc} f^{ade} \frac{1}{\partial^+}\left(\partial^+A^b\bar{A}^c\right)\frac{1}{\partial^+}\left(\partial^+\bar{A}^d A^e\right) \nn\\
&&\hspace{-0.7 cm}+\,\frac{i}{\sqrt{2}}\bar{\chi}^a\left(\frac{\Box}{\del^+}\delta^{ac}-2gf^{abc}\frac{1}{\del^+}(\partial\bar{A^b}+\bar{\partial}A^b) +2gf^{abc}{\bar{A}^b}\frac{\partial}{\del^+}\right)\chi^c + i\sqrt{2} gf^{abc} \bar{\chi}^a\frac{\bar{\partial}}{\del^+}(A^b \chi^c)\nn\\
&&\hspace{ -0.5 cm}\,+\,i\sqrt{2}\,g^2 f^{abc}f^{bde}\,\bar{\chi}^a\frac{1}{\del^{+2}}(A^d\del^+\bar{A}^e+\bar{A}^d\del^+{A}^e)\chi^c-i\sqrt{2}\,g^2 f^{abd}f^{bec}\bar{\chi}^a \bar{A}^d\frac{1}{\del^+}(A^e\chi^c)\nn\\
&&\hspace{1cm}+g^2f^{abc}f^{ade}\frac{1}{\del^+}(\bar{\chi}^b\chi^c)\frac{1}{\del^+}(\bar{\chi}^d\chi^e)\,.\nn\\
\eea

\section{Quartic fermion term}

For simplicity we study the $d=4$ case. The analysis however holds true in any $d$. The path integral takes the form
\bea
Z=\int DA^a D\bar{A}^a D\chi ^a D\bar{\chi}^a\,\, \text{exp}\left[i\int d^4x\,\mathcal{L}_1+B\right]\,,
\eea
\ndt where $\mathcal{L}_1$ contains all terms in (\ref{Final L}) except the four fermion interaction term and $B$ is the four fermionic interaction. We now expand the exponent of $B$ to the linear order (since we are working to order $g^2$ and $B$ is exactly of that order)

\bea
Z=\int DA^a D\bar{A}^a D\chi ^a D\bar{\chi}^a\,\,(1+i\int d^4x\,B)\, \text{exp}\left[i\int d^4x\,\mathcal{L}_1\right]\,,
\eea
\bea
Z=Z_0+Z_1\,.
\eea
\ndt The fermion determinant can be evaluated in the term $Z_0$. Let us denote this determinant by $\Delta_F(A,\bar{A},g)$. The form of $Z_0$ is then
\bea
Z_0=\int DA^a D\bar{A}^a \,\,\Delta_F(A,\bar{A},g) \,\,\exp\left[i\int d^4x\,\mathcal{L}_{YM}\right]\,,
\eea

\ndt where $\mathcal{L}_{YM}$ denotes the Yang-Mills Lagrangian. The term $Z_1$ is the path integral $Z_0$ with the four fermion interaction term as an insertion. The term $Z_1$ may be computed using the standard technique of introducing sources by considering

\bea
Z_0[J]&=&\int DA^a D\bar{A}^a D\chi ^a D\bar{\chi}^a\,\nn\\
&& \text{exp}\left[i\int d^4x\,\mathcal{L}_{YM}+\frac{i}{\sqrt{2}}\left((\bar{\chi}^a+\bar{J}^b Q^{-1\,ba})Q^{ac}(\chi^c+Q^{-1\,cd}J^d)-\bar{J}^a Q^{-1\,ac}J^c\right)\right]\,.\nn\\
\eea

\ndt Here $Q^{ac}$ denotes the quadratic operator in the fermionic part of the Lagrangian while $Q^{-1\,ac}$ denotes the fermion propagator in the presence of gauge field. A change of variables in the path integral from $\chi^a$ to $\chi^{'\,a}=\chi^a+Q^{-1\,ac}J^c$ and similarly for its complex conjugate allows us to integrate the fermion fields and get a factor of $\Delta_F(A,\bar{A},g)$. By differentiating $Z_0[J]$ with respect to the sources and putting them to zero we find the form of $Z_1$ to be

\bea
Z_1=\int DA^a D\bar{A}^a\,\,\Delta_F(A,\bar{A},g)G_4(A,\bar{A},g)\,\,\exp\left[i\int d^4x\,\mathcal{L}_{YM}\right]\,.
\eea
Thus,
\bea
Z=\int DA^a D\bar{A}^a\,\,\Delta_F(A,\bar{A},g)[1+G_4(A,\bar{A},g)]\,\,\exp\left[i\int d^4x\,\mathcal{L}_{YM}\right]\,.
\eea

\ndt We can define $\Delta'_F(A,\bar{A},g)=\Delta_F(A,\bar{A},g)[1+G_4(A,\bar{A},g)]$ as the effective fermion determinant. Note that $G_4$ is a product of two fermion propagators. The fermion propagator in the presence of a gauge field is an infinite series in the coupling $g$. Since $Z_1$ is itself at order $g^2$, we must only consider the leading term in the series which is nothing but the free fermion propagator given by $\del^+C(x-y)$. But since all the four fermion fields are at the same space-time point, $G_4$ vanishes as $\del^+C(0)=0$.\\

\ndt Thus four fermion term  starts contributing to the fermion determinant from order $g^4$.

\end{document}